\documentclass[aps,prl,floats,floatfix,twocolumn]{revtex4}

\usepackage{ifthen}
\usepackage{ifpdf}
\usepackage{color}

\ifpdf
\usepackage{graphicx}
\usepackage{epstopdf}
\else
\usepackage{graphicx}
\usepackage{epsfig}
\fi

\usepackage{latexsym}
\usepackage{amsmath}
\usepackage{amssymb}
\usepackage{bm}
\usepackage{wasysym}

\newcommand{\eexp}{\mbox{e}^}

\newcommand{\mylabel}[1]{\label{#1}} 
\newcommand{\beq}{\begin{eqnarray}}
\newcommand{\eeq}{\end{eqnarray}} 
\newcommand{\be}[1]{\begin{eqnarray}\ifthenelse{#1=-1}
{\nonumber}{\ifthenelse{#1=0}{}{\mylabel{e#1}}}}
\newcommand{\ee}{\end{eqnarray}} 

\newcommand{\Eq}[1]{\textcolor{blue}{Eq.\!\!~(\ref{#1})}} 
\newcommand{\Fig}[1]{\textcolor{blue}{Fig.}\!\!~\ref{#1}} 
\newcommand{\hide}[1]{}
\newcommand{\rmrk}[1]{#1}   
\renewcommand{\cite}[1]{\textcolor{blue}{[\onlinecite{#1}}]} 

\begin{document}

\title{Minimal Fokker-Planck theory for the thermalization of mesoscopic subsystems}

\author{Igor Tikhonenkov$^1$, Amichay Vardi$^1$, James R. Anglin$^2$, and Doron Cohen$^3$}

\affiliation{
\mbox{$^1$Department of Chemistry, Ben Gurion University of the Negev, Beer Sheva 84105, Israel}
\mbox{$^2$OPTIMAS Research Center and Fachbereich Physik, Technische Universit\"at Kaiserslautern, D-67653 Kaiserslautern, Germany}
\mbox{$^3$Department of Physics, Ben Gurion University of the Negev, Beer Sheva 84105, Israel}
} 

\begin{abstract}
We explore a minimal paradigm for thermalization, consisting of two weakly-coupled, low dimensional, non-integrable subsystems. As demonstrated for Bose-Hubbard trimers, chaotic ergodicity results in a diffusive response of each subsystem, insensitive to the details of the drive exerted on it by the other. This supports the hypothesis that thermalization can be described by a Fokker Plank equation.  We also observe, however, that Levy-flight type anomalies may arise in mesoscopic systems, due to the 
wide range of time scales that characterize `sticky' dynamics.  
\end{abstract}

\maketitle


The emergence of irreversibility from reversible Hamiltonian mechanics remains an open fundamental question, even after a century of effort. Recent advances in computational as well as experimental technique may at last bring answers within reach. The biggest challenge of this quest is the sheer technical difficulty of solving the Hamiltonian evolution of quantum many-body systems, even when they are quite small and isolated. In this Letter we propose to leap over a significant barrier of understanding, by using Hamiltonian results from a tractable but non-trivial system, to support an extension of an established phenomenological theory, into a substantially more challenging regime. The result we thereby derive is a simple theory that can then both guide, and be tested by, subsequent numerical investigations, as well as currently feasible experiments.

We address the thermalization of two nonlinear Hamiltonian subsystems 
that are weakly coupled together, where the combined system is isolated and undriven. 
The equilibration of such subsystems is postulated in the Zeroth Law of Thermodynamics, 
reflecting the assumption that microscopic dynamics is unobservably fast, while slower
macroscopic dynamics remains nontrivial. Accordingly, weakly 
coupled subsystems, each having strong internal interactions, provide the minimal 
paradigm for the emergence of thermodynamics from closed-system mechanics.

Following the Fermi-Pasta-Ulam numerical experiment, most studies of dynamical equilibration have historically focused on large, extended systems \cite{extended1,extended2}, where the treatment of even one strongly interacting system is 
quite impossible in microscopic detail. With experimental access to controlled mesoscopic systems, 
attention has more recently been drawn to thermalization phenomena in small systems, 
taking into account dynamical chaos \cite{supp,Dorfman} 
and quantum effects \cite{qthermalization1,qthermalization2,qthermalization3,fpe}.
The traditional analysis of thermalization has nonetheless largely remained within 
the assumptions inherited from the macroscopic problem. It is common to assume that 
at least one of the two coupled systems is ``big", and hence can be drastically approximated,
either as a phenomenologically described reservoir, or as a time-dependent external parameter. 
The present Letter is motivated by the realization that the study of isolated thermalization 
of two subsystems is no longer so unthinkably intractable. 
It is merely extremely difficult. Our proposal is to leverage our understanding of driven
chaotic systems to overcome this difficulty, by viewing each subsystem as driving the other.

{\bf The statistical approach.-- }
The statistical description of driven chaotic systems by means 
of a Fokker-Planck equation (FPE) for their energy distribution 
\cite{Wilk,Jar,frc,WilkA,rsp,kafriN} is based on the ergodic adiabatic 
theorem \cite{Ott}. Quantum and classical systems can be embraced in a unified notation 
by writing the energy $\varepsilon$ as a function of the phase space volume~$n$ 
of the constant-energy hypersurface. The density of states is $g(\varepsilon) = dn/d\varepsilon$, 
and the micro-canonical inverse temperature is $\beta(\varepsilon) = d\ln(g)/d\varepsilon$.
Upon quantization, the Wigner-Weyl formalism implies that~$n$ 
corresponds to the discrete index of the energy levels $\varepsilon_n$, 
and if these levels are dense enough, they can be approximated as a quasi-continuum.
One then makes a coarse-grained description of the slow evolution of the system, 
and derives an FPE to describe the evolution of the time-dependent energy probability 
distribution $\rho(\varepsilon,t)$.

{\bf FPE for a driven system.-- }
If a chaotic system is driven weakly, its energy changes slowly, 
and $\rho(\varepsilon,t)$ obeys a probability-conserving FPE, 
whose diffusion term has a coefficient~$D$, 
proportional to the strength of the driving \cite{Wilk,Jar,frc,kafriN}. 
By Liouville's theorem, a distribution $\rho(\varepsilon) \propto g(\epsilon)$ 
should be a time-independent solution of the FPE.
Hence it is deduced that the drift term in the FPE is universally 
related to~$D$, and the complete phenomenological equation is established.

{\bf FPE for coupled subsystems.-- }
We now extend the single-system FPE phenomenology \cite{Wilk,Jar,frc}
to the case of thermalization of  {\em two}  subsystems. 
Each subsystem (${i=1,2}$) is characterized by its density 
of states $g_{i}(\varepsilon_{i})$, and by its microcanonical 
inverse temperature $\beta_{i}$. 
Thanks to conservation of energy the thermalization is within subspaces of constant 
energy $\varepsilon_1(n_1)+ \varepsilon_2(n_2) = \mathcal{E}$.
Accordingly we set $\varepsilon_{1} = \varepsilon$, 
and $\varepsilon_{2} = \mathcal{E}-\varepsilon$, 
and construct an FPE for the probability density $\rho(\varepsilon,t)$ 
that describes how the energy is divided between the two subsystems.

It again follows from Liouville's theorem that an ergodic distribution 
$\rho(\varepsilon) \propto g(\varepsilon)\equiv g_{1}(\varepsilon)g_{2}(\mathcal{E}-\varepsilon)$,  
should be a stationary solution. This fixes the form 
of the FPE, and implies the functional form of the drift term:  
\beq\label{FPE1}
\frac{\partial \rho}{\partial t} \ &=& \ 
\frac{\partial}{\partial \varepsilon}
\left(g(\varepsilon)D(\varepsilon) \frac{\partial}{\partial \varepsilon}
\left(\frac{1}{g(\varepsilon)}\rho\right)\right)
\\ \label{FPE2}
\ &=& \ 
-\frac{\partial}{\partial \varepsilon}
\left(
A(\varepsilon) \rho -  
\frac{\partial}{\partial \varepsilon}\left[D(\varepsilon)\rho\right]
\right)
\eeq
It is important to notice that the diffusion coefficient $D$ may depend on $\varepsilon$.
The optional way \Eq{FPE2} of writing this FPE demonstrates that the `drift velocity' $A$
is related to the diffusion as follows:  
\beq\label{Adef}
A(\varepsilon) \ \  = \ \ \partial_{\varepsilon} D + (\beta_1-\beta_2) D\;.
\eeq
To see more clearly the connection of \Eq{Adef} with traditional thermodynamics, 
assume that each of the subsystems is prepared independently in a canonical state, 
with temperature $T_i$, 
such that $g_i(\varepsilon_i) \exp(-\varepsilon_i/T_i)$ describes its energy distribution.
Integrating both sides of \Eq{Adef} with this probability measure, 
and integrating by parts the first term on the right, 
one obtains \cite{supp} a mesoscopic Einstein relation, 
like those previously derived~\cite{kbb,kafri} 
using master-equation or fluctuation-theorem approaches:
\beq\label{Einstein}
\frac{d}{dt}\langle\varepsilon\rangle 
\ = \ \langle A(\varepsilon) \rangle  
\ = \ \left( \frac{1}{T_1}-\frac{1}{T_2} \right) \langle D \rangle \;,
\eeq
This result offers insight into the distinct behaviors 
of micro-canonical energy fluctuations and canonical averages.
The canonical version \Eq{Einstein} implies that energy always 
flows from the higher to the lower canonical temperature, 
but the more general mesoscopic version \Eq{Adef} 
implies that energy flow is not necessarily from the  
higher to the lower micro-canonical temperature, 
and may depend on the functional form of $D(\varepsilon)$. 
This is not in contradiction with the Zeroth Law of 
thermodynamics: energy fluctuates between finite systems in equilibrium, 
such that their average micro-canonical temperatures need not be equal. 
The ergodic solution $\rho \propto g(\varepsilon)$ 
around which \Eq{FPE1} has been constructed implies
only that the most probable $\varepsilon$ is the one for which 
$\beta_{1}(\varepsilon)=\beta_{2}(\mathcal{E}-\varepsilon)$.

{\bf Fluctuation-dissipation phenomenology.-- }
The derivation of \Eq{FPE1} is phenomenological, but based on simple assumptions 
that can be tested. Since these include weak coupling, it is further consistent 
to compute $D$ using the Kubo formula.
Writing the interaction as $\mathcal{H}=Q^{(1)}Q^{(2)}$, 
and defining $\tilde{S}^{(i)}(\omega)$ as the power spectrum
of the fluctuating variable $Q^{(i)}(t)$, it reads \cite{Kubo}
%
%
\beq\label{KuboD}
D \ \ = \ \ \int_0^{\infty} \frac{d\omega}{2\pi} \, \omega^2  \, \tilde{S}^{(1)}(\omega) \, \tilde{S}^{(2)}(\omega)
\eeq
With this addition, the FPE phenomenology provides a generalized fluctuation-dissipation relation 
that connects the systematic energy flow between the subsystems with the intensity of the fluctuations.

{\bf Reasoning.-- }
\rmrk{
When two undriven subsystems are coupled 
to each other, the effect of one 
subsystem (call it "agent") 
on the other (call it "system")
is like that of driving. 
For the purpose of obtaining \Eq{FPE1} we have 
assumed that the interaction results in diffusion 
that can be calculated using \Eq{KuboD}.  
Future studies of coupled systems must test 
this assumption in full, but one key point 
remains to be established with regard to 
the driven single-subsystem dynamics: 
The agent-system interaction will typically 
couple many quantum levels, 
even if it is weak in classical terms;
In such strongly non-adiabatic circumstances, 
energy diffusion, and the applicability of \Eq{KuboD},
have not been demonstrated \cite{supp}. 
Below we complete this paper by a numerical demonstration 
that highlights the role of chaos in obtaining diffusive 
dynamics for a driven subsystem, supporting the feasibility 
of the above reasoning for experimentally relevant systems.
}

{\bf Testing ground.-- }
Few-mode Bose-Hubbard systems are a promising testing ground, 
since they are experimentally accessible and highly tunable \cite{BHH}, 
and theoretically tractable by a wide range of techniques. 
Since boson number is conserved, their Hilbert spaces are of finite dimension, 
and yet their classical dynamics can be non-integrable.  
The smallest Bose-Hubbard system admitting chaos without 
external driving is the three-mode {\em trimer}~\cite{trimer1,trimer2,trimer3,trimer4,trimer5,trimerD1,trimerD2}, 
described by the Bose-Hubbard Hamiltonian (BHH):
\be{1}
\mathcal{H} = 
\frac{K}{2}\sum_{i=1,2} \left(a_i^{\dag} a_0 + a_0^{\dag}a_i\right) 
+ \frac{U}{2}\sum_{i=0,1,2} a_i^{\dag} a_i^{\dag} a_i a_i  ~,
\eeq
Here $i=0,1,2$ label the three modes, $a_{i}$ and $a_{i}^{\dagger}$ are canonical destruction and creation
operators in second quantization, $K$ is the hopping frequency, and $U$ is the on-site interaction. 
The Hamiltonian $\mathcal{H}$ commutes with the total particle number $\mathcal{N}=\sum_{i}a^{\dagger}_{i}a_{i}$, 
and hence, without loss of generality, we regard $\mathcal{N}$ as having a definite value~$N$. 
Driving is then implemented by setting ${K = K_0 + K_d\sin(\Omega t)}$. 
Consequently the total Hamiltonian has the structure $\mathcal{H}_0+f(t)W$, 
where the perturbation operator $W$ is identified as the first sum in \Eq{e1}, 
and the driving field is $f(t)=(K_d/2)\sin(\Omega t)$.

\begin{figure}[t]
\includegraphics[width=\hsize]{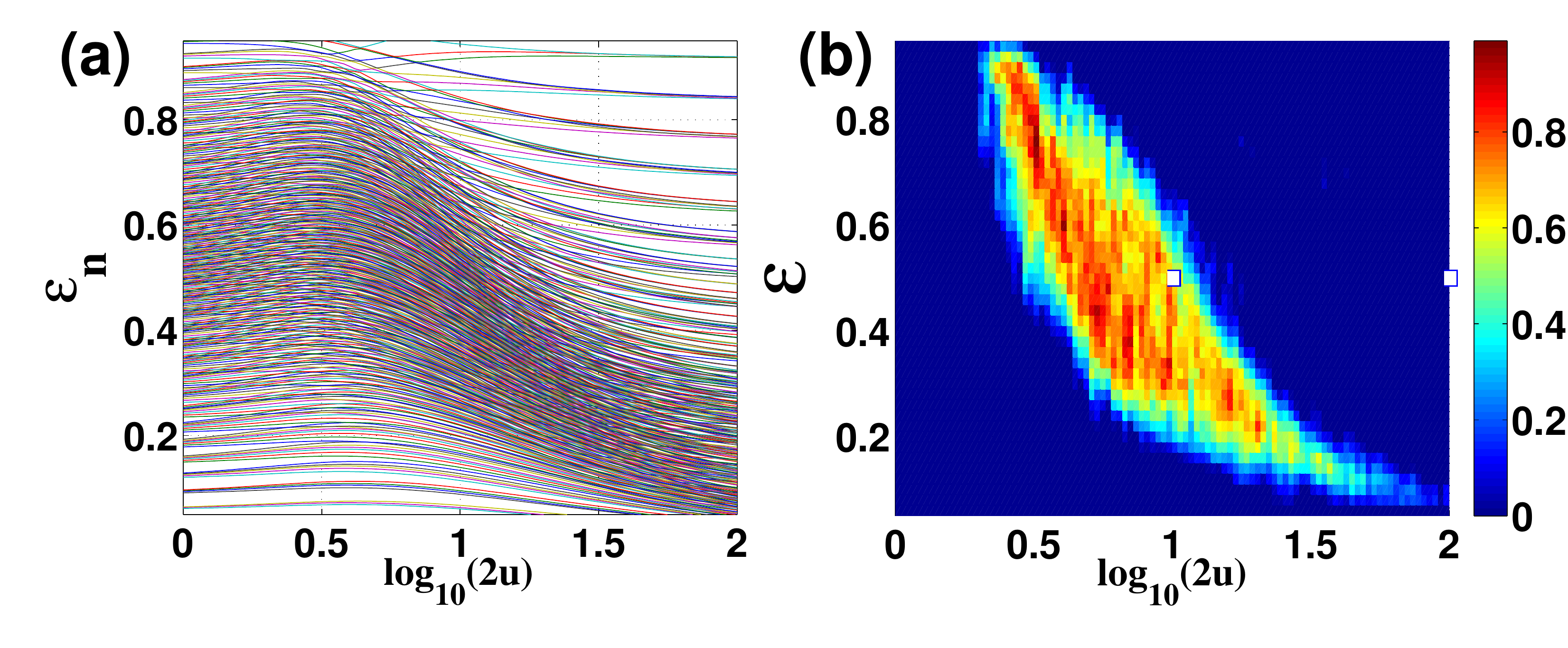}\\

\caption{
The energy spectrum of the \rmrk{unperturbed BHH}.
In panel (a) the scaled eigen-energies $\varepsilon_n$ \rmrk{of $\mathcal{H}_0$}
are plotted versus the scaled interaction parameter $u$, 
for $N=35$ particles.
%
%
The level spacing statistics is characterized by the Brody 
parameter (${0<q<1}$), which is displayed 
in Panel (b) for a system with ${N=120}$ particles.  
In the energy range where the motion is chaotic ${q\sim1}$.
Square symbols indicate the preparations that were 
used for the simulations in \Fig{f3}.
}

\label{f1}
\end{figure}

{\bf Chaoticity.-- } 
The underlying classical dynamics
is defined~\cite{supp} by replacing the operators $a_{i}$
in the Heisenberg equations of motion \cite{trimerE}
with complex c-numbers $\sqrt{n_i}\eexp{i\varphi_i}$.
In the absence of driving, up to trivial rescaling, 
the classical equations depend only 
on the single dimensionless parameter ${u=NU/K_{0}}$. 
The chaoticity of the motion that is generated by $\mathcal{H}_0$
is reflected in the local level statistics, 
and can be quantified by the Brody parameter $0{<}q{<}1$ \cite{Brody81},
such that $q{=}0$ indicates a Poissonian level-spacing distribution 
(characteristic of integrable dynamics), 
while higher values indicate the approach to Wigner level-spacing distribution 
(indicating chaotic dynamics). 

\Fig{f1}a displays the spectrum $\varepsilon_n$, obtained by numerical 
diagonalization of $\mathcal{H}_0$, as a function of $u$. 
For graphical presentation we shift and scale the energy spectrum, 
for each $u$, into the same range ${\varepsilon\in[0,1]}$,
such that ${\varepsilon=0}$ and ${\varepsilon=1}$ are the 
ground energy $E_0$ and the highest energy $E_{\text{max}}$ respectively.
By plotting $q$ vs ${(u,\varepsilon)}$, as in \Fig{f1}b, we can identify 
the $\varepsilon$ range within which the motion is chaotic at any given value of $u$. 
See \cite{supp} for technical details.
We have verified the implied chaoticity by plotting representative  
classical Poincare cross-sections.

In the numerical simulations we consider an ${N=50}$ particle 
system with two representative values of~$u$.
The case ${u=5}$, for which there is a wide chaotic range ${0.2<\varepsilon<0.6}$, 
is contrasted with $u=50$, for which the motion is globally 
quasi-integrable due to self-trapping.

\begin{figure}[t!]
\includegraphics[width=\hsize]{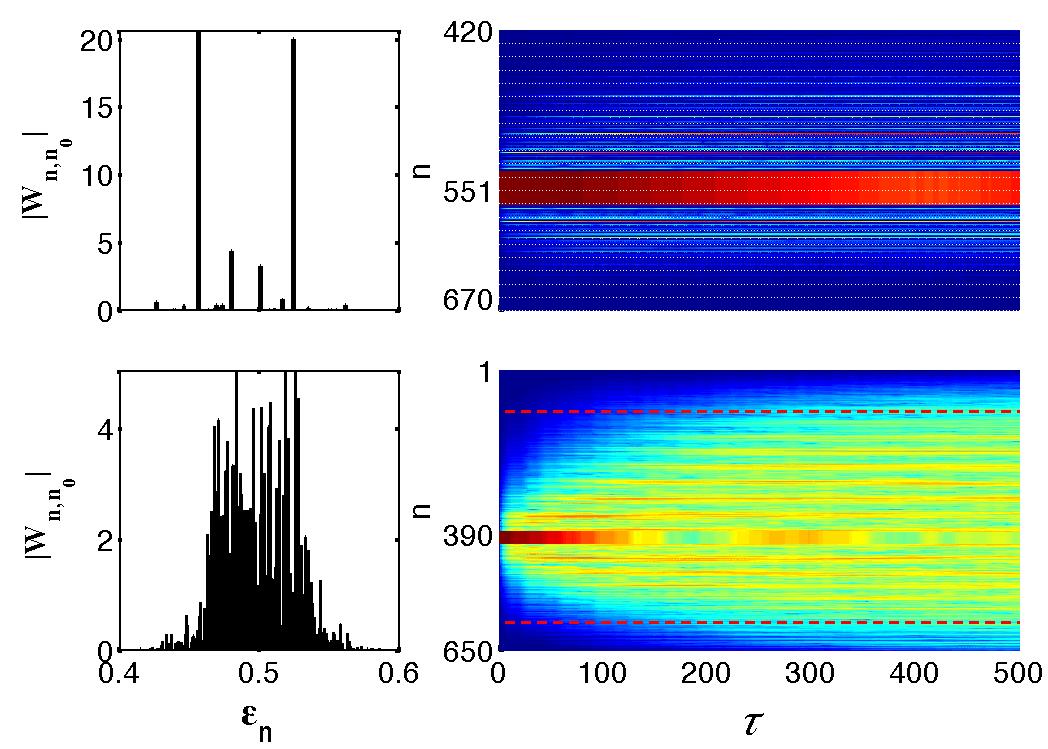}

\caption{
The quantum probability distribution $p_n(t)$ 
for representative simulations is imaged 
as a function of time (right). The short-time 
energy-spreading profile is determined 
by the perturbation matrix $|W_{n,n_0}|^2$ (left).
The number of particles is ${N=50}$.
The upper set is for $u=50$, and the lower is for $u=5$. 
The strength of the driving is ${K_d/K_0=0.1}$.
For the time axis we use dimensionless 
units ${\tau = (E_{\text{max}}-E_{0})t/\hbar}$, 
and the scaled driving frequency in both 
simulations is $\Omega\approx 0.03$.  
%
%
The image of the initial level is vertically zoomed, 
and it has the energy ${\varepsilon \approx 0.5}$.     
The boundaries of the chaotic sea in the
lower image are indicated by the horizontal dashed lines.
}
\label{f3}
\end{figure}

{\bf Energy diffusion.-- }
In the absence of driving the energy is a constant of motion. 
Driving induces transitions between energy eigenstates, 
leading to a time-dependent spread in energy $\Delta\varepsilon(t)$. 
This dispersion is defined as the square root of the variance Var$(\epsilon_n)$ 
that is associated with the probability distribution 
\beq
p_n(t) \ \ = \ \ \Big|\langle \varepsilon_n | \Psi(t)\rangle\Big|^2~.  
\eeq
In \Fig{f3} we plot the time evolution of the quantum energy distribution $p_n(t)$ 
in response to driving that is quantum mechanically 
large (many levels are mixed) but classically small (${K_d\ll K_0}$).  
We contrast the response in the chaotic ($u=5$) 
and in the quasi-integrable ($u=50$) regimes.  
Dramatic differences are observed. In both cases, the energy 
distribution in the very early stages of the evolution reflects the 
band profile of the perturbation matrix $W_{n,n_0}$, 
\rmrk{where $n_0$ is the initial level}, 
as expected from time-dependent first-order perturbation theory.  
Later in the evolution higher orders of perturbation theory dominate. 
This leads in the quasi-integrable case to Rabi-like oscillations that 
have no relation to the classical dynamics.  But in the chaotic regime 
one observes that the driving is capable of inducing diffusive-like energy spreading. 
This diffusive spreading is restricted to the chaotic energy window,
and features remarkable correspondence with the classical simulation.

\begin{figure}
\includegraphics[width=\hsize]{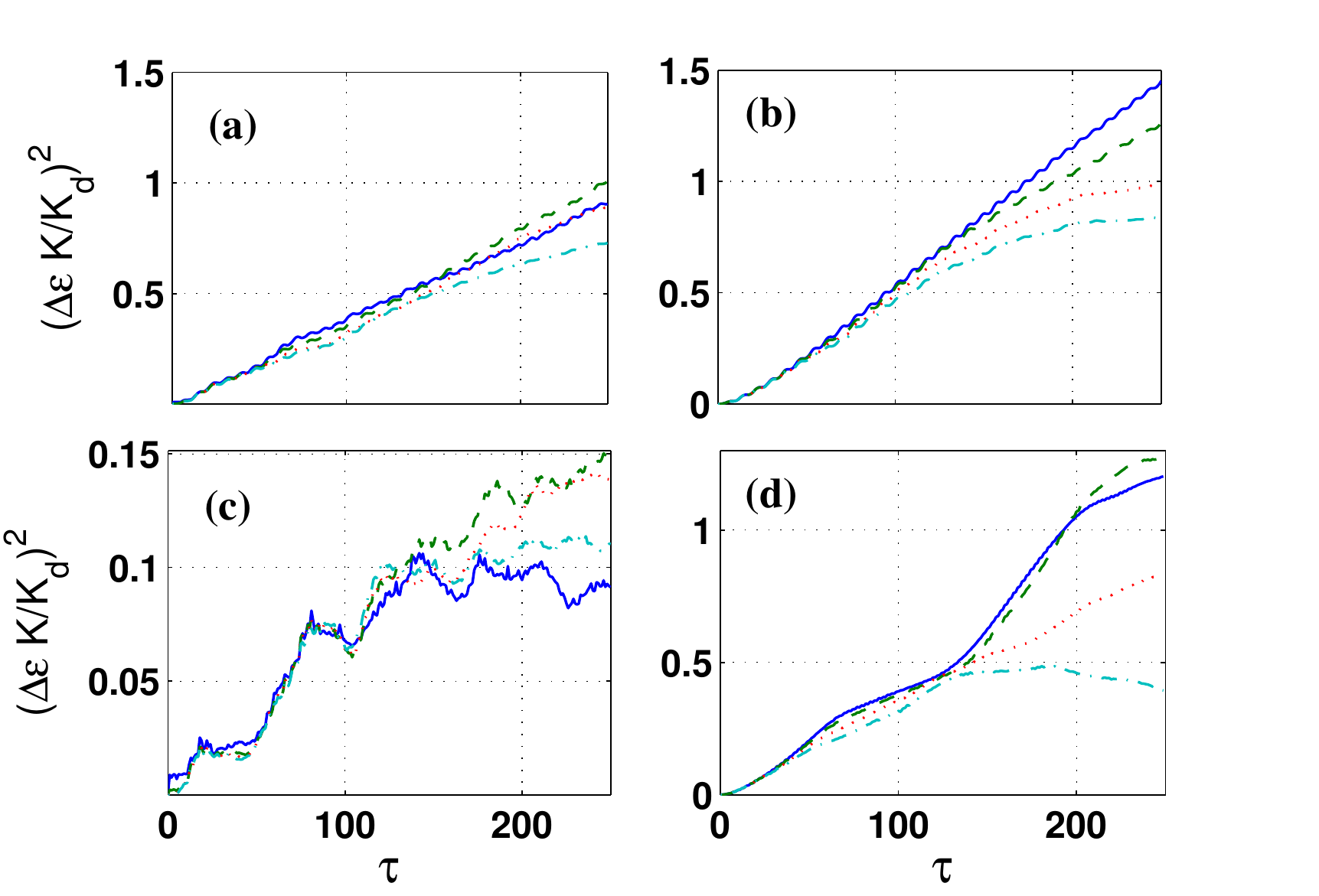}

\caption{
The scaled variance as a function of the scaled time 
in the classical (left) and in the quantum (right) 
simulations. The value of the scaled interaction 
parameter is $u=5$ (upper panels) and $u=50$ (lower panels).
Note the different scale of the vertical axis 
for the quantum vs classical simulations in panels~(c) and~(d). 
Clearly quantum-to-classical correspondence fails 
in the quasi-integrable regime. The values of $K_d/K_0$  
are $0.025$ (blue), $0.05$ (green), $0.075$ (red), and $0.1$ (cyan). 
The driving frequency is as in \Fig{f3}. 
}

\label{f4}
\end{figure}

{\bf Linear response.-- }
The diffusive energy spreading in the chaotic regime can be quantified by the time evolution 
of the energy variance. In \Fig{f4} we plot the time evolution 
of $\Delta\varepsilon$ for both the chaotic and the integrable cases. In both cases we compare the dispersion obtained under the classical equations of motion for the driven system, starting from a micro-canonical ensemble, to that obtained from quantum evolution from an eigenstate with the same energy. As anticipated for diffusive energy spreading, we observe that in the chaotic regime $(\Delta\varepsilon)^2 \approx 2Dt$ with diffusion coefficient ${D\propto K_d^2}$, as assumed in the Kubo linear response formula \Eq{KuboD}.  

\begin{figure}
\includegraphics[width=\hsize]{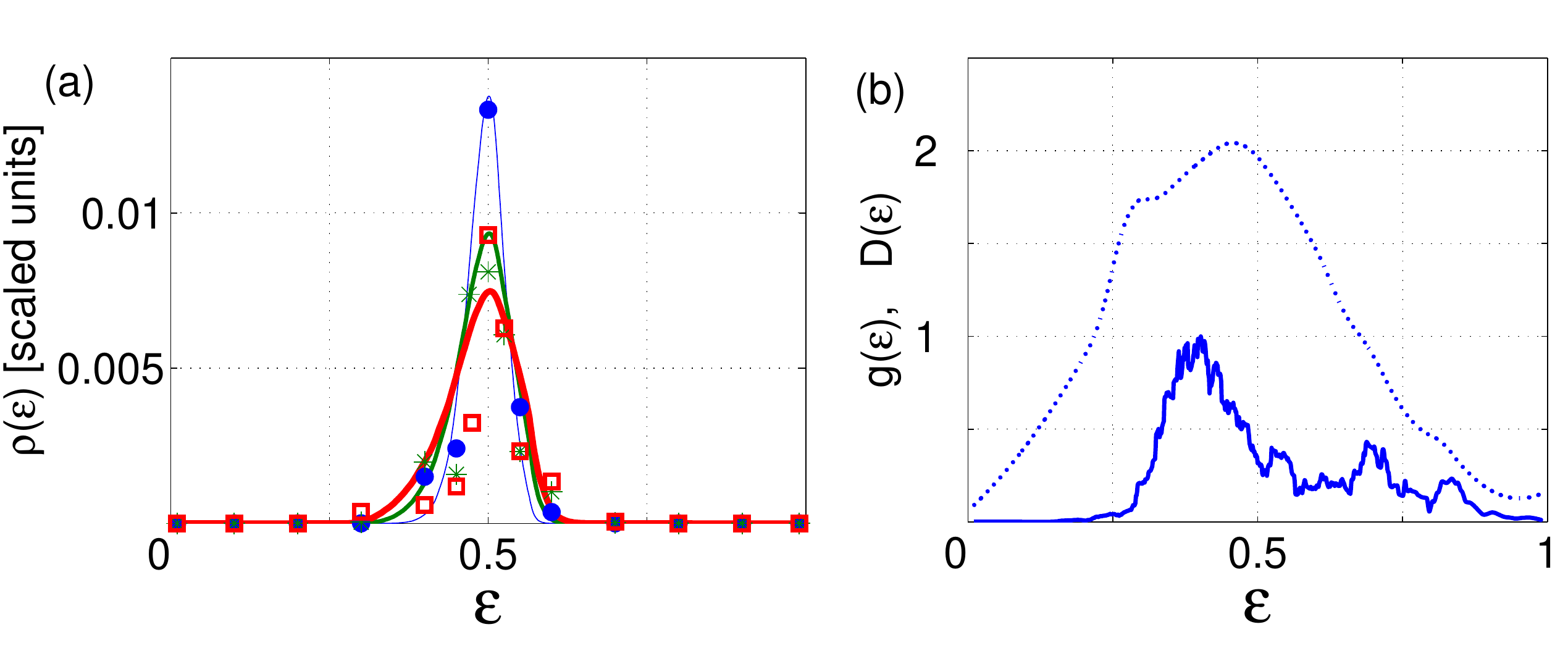}

\caption{
(a) The evolving spreading profile $\rho(E)$,  
referring to the simulation that has been imaged 
in the lower panel of \Fig{f3}.
The solid lines (calculation) 
and the associated symbols (simulation) 
are for $\tau=149$ (narrower),
and  $\tau=299$ (wider), 
and $\tau=448$ (widest). 
The lines are based on the numerical solution of \Eq{FPE1}. 
(b) The density of states $g(E)$ (dotted line) 
and the diffusion coefficient $D(E)$ (solid line)
were deduced from the diagonalization  
of the BHH and \Eq{KuboD}, 
see \cite{supp} for technical details. 
}

\label{f4PE}
\end{figure}

\rmrk{
In \Fig{f4PE} we compare the diffusive energy distribution that is observed in \Fig{f3},   
with the solution of the FPE \Eq{FPE1}, using the diffusion coefficient from  \Eq{KuboD}, 
see \cite{supp} for technical details. 
The agreement is good, confirming that weakly driven chaotic quantum systems 
can indeed exhibit energy diffusion in regimes realistic for experimental Bose-Hubbard systems, 
and solidifying the basis of our phenomenological argument for the FPE description 
of inter-subsystem equilibration.}

\begin{figure}[t!]
\includegraphics[width=\hsize]{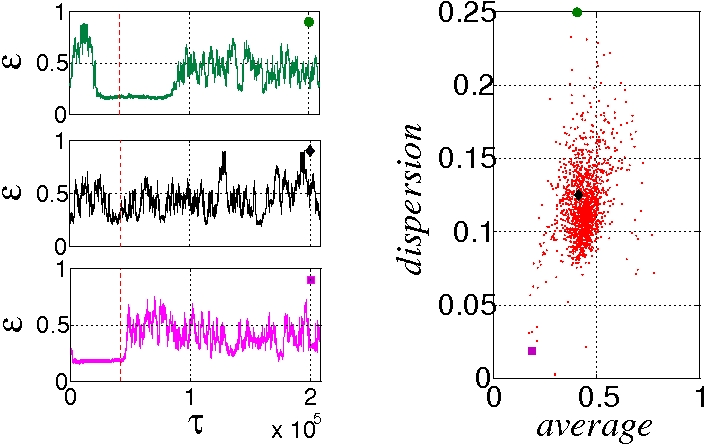}

\caption{
On the left the scaled energy $\varepsilon(t)$ 
as a function of time is plotted for a few representative 
trajectories. The right panel displays the average 
value and the dispersion of $\varepsilon$ 
within the time interval ${0<\tau<40000}$, 
for the representative trajectories (symbols), 
as well as for many other trajectories (points).
Low dispersion values reflect the finite probability 
to encounter sticky motion.  
The parameters are the same 
as in the lower panel of \Fig{f3}, 
with initial points that have 
the energy ${\varepsilon = 0.3}$. 
}

\label{f5}
\end{figure}

{\bf Multiple timescales.-- }
Having established quite good quantum-to-classical correspondence 
in the chaotic regime, one wonders whether classical dynamics may 
indicate features that go beyond simple energy diffusion.
\Fig{f5} illustrates the {\em classical} time dependence of $\varepsilon(t)$, 
and characterizes it by its average value and dispersion.
Since the phase space has dimension greater than two, the possibility of Arnold diffusion 
guarantees that the motion is ergodic within the chaotic sea. 
This means that we can regard different trajectory segments 
as uncorrelated pieces of the same infinite time trajectory. 
If we had ergodic motion with a well defined characteristic 
time, all the segments would have the same average and dispersion. 
But this is not what we see: the segments have large variation 
in their dispersion, since they do not uniformly fill the whole 
chaotic region. Rather, the trajectories contain episodes with long 
dwell times within some sticky regions in phase space, whose
existence has been confirmed with a Poincar\'e section, see \cite{supp}.
If we had an unlimited computation power, obviously the 
expectation is to have coincidence of all the points 
in the right panel of \Fig{f5}. But in practice we can address 
only finite time intervals, and therefore the points are scattered 
over a large range.

{\bf Discussion.-- }
We have considered few-mode Bose-Hubbard systems as tunably
chaotic systems, which in chaotic regimes respond generically to weak
driving with energy diffusion at a rate proportional to the square
of the driving strength, $K_d^2$. Consequently we deduce that 
the thermalization of coupled Bose-Hubbard sub-systems can plausibly 
be described by a phenomenological FPE, namely \Eq{FPE1}. 

We have also obtained some insight 
on how such phenomenological theories are affected by taking 
into account specific semi-classical features of the dynamics.
A~small sub-system can exhibit multiple time scales in its equilibration,
because its phase-space contains sticky regions with long dwell times. 
In principle this may give rise to non-Gaussian features, 
and L\'evy-flight related deviations from strict diffusive behavior. 
Even if these possibilities do not manifest strongly in energy spreading
of small, driven systems, because the explored phase space volume is small,
they may perhaps become important in multi-component composite systems.
It would be important to recognize, then, in interpreting experiments 
or simulations intended to test \Eq{FPE1}, that some deviations 
from its diffusive assumptions may not represent errors in its depiction 
of the mesoscopic onset of irreversibility, 
but only the fading traces of microscopic behavior.


{\bf Acknowledgments.-- }
This research was supported by the Israel Science Foundation (grant Nos. 346/11 and 29/11) 
and by the United States-Israel Binational Science Foundation (BSF).



\clearpage

\onecolumngrid

\begin{center}
{\LARGE Supplementary material}  
\end{center}

\ \\

\twocolumngrid

{\bf The role of chaos.-- }
The key obstacle for thermalization can be appreciated by 
considering the common paradigm for driven {\em integrable} system:
the so called ``kicked rotor" as described by the ``standard map" \cite{qkr}.
In the absence of driving the system 
is integrable. The driving amplitude is $K$. Below a critical 
value $K_c\approx0.97$ there is no diffusion in energy 
due to Kolmogorov-Arnold-Moser blocking.
For somewhat larger values ${D\propto (K-K_c)^3}$. 
Only for strong driving amplitude one observes 
a quasi-linear dependence ${D\propto K_d^2}$.  

The dependence of $D$ on the driving amplitude $K$ is 
strikingly different in the case of a driven {\em chaotic} system: 
Following the ergodic adiabatic theorem of \cite{Ott},  
it has been realized \cite{Wilk,Jar,frc} that a linear response 
dependence ${D\propto K^2}$ shows up for arbitrarily small 
driving amplitude with arbitrarily small driving frequency (``DC limit"). 

In the quantum domain the applicability of linear response 
has been first challenged \cite{WilkA} and later re-analyzed 
and established \cite{rsp} using a random matrix theory (RMT)
and semi-classical perspectives. The existence of the  
underlying classical dynamics is essential in order 
to avoid RMT anomalies that arise beyond 
the regime of 1st order perturbation theory \cite{crs}. 

Strangely enough the semi-classical implied robustness 
of the quantum diffusive behavior has never been verified, 
to the best of our knowledge, for a realistic quantized system.
More precisely - there are numerous simulations in the 
quantum adiabatic regime where the transitions are mainly 
between neighboring levels. But the regime of our interest is 
different: our interest is in driving intensities 
that can be regarded as quantum mechanically large, 
but still semi-classically small. This is the regime 
where the energy landscape can be regarded as a quasi-continuum 
and quantum-to-classical correspondence can be expected.    

\onecolumngrid

\ \\

{\bf Obtaining the Einstein relation.-- }
Let us see how \Eq{Einstein} is obtained from \Eq{Adef}.
We assume that both systems are independently 
in a canonical state. Accordingly the joint probability 
distribution is 
\beq
\rho(\varepsilon_1,\varepsilon_2) \ = \
\frac{1}{Z_1 Z_2} g_1(\varepsilon_1) g_2(\varepsilon_2) 
\exp\left[-\frac{\varepsilon_1}{T_1}  -\frac{\varepsilon_2}{T_2}\right]
\eeq
Averaging over the drift term we get 
the rate of energy absorption:
\beq 
\langle A(\varepsilon)\rangle  =  
\frac{1}{Z_1 Z_2}
\int d\mathcal{E} \exp\left[-\frac{\mathcal{E}}{T_2}\right] 
\int d\epsilon 
\ \left[ \partial_{\varepsilon} D + (\beta_1-\beta_2)D \right] \ 
g_1(\varepsilon)g_2(\mathcal{E}-\varepsilon) 
\ \exp\left[-\left(\frac{1}{T_1}-\frac{1}{T_2}\right)\varepsilon\right]
\eeq
Doing integration by parts on the term that involves $\partial_{\varepsilon} D$, 
and noting that $\partial_{\varepsilon}g_{1,2} =  \pm g_{1,2}\beta_{1,2}$ 
is a contribution that cancels with the $(\beta_1-\beta_2)D$ term,  
one observes that we are left with 
\beq \nonumber
\langle A(\varepsilon)\rangle   =   
\frac{1}{Z_1 Z_2}
\int d\mathcal{E} \exp\left[-\frac{\mathcal{E}}{T_2}\right] 
\int d\epsilon 
\ D \ g_1(\varepsilon)g_2(\mathcal{E}-\varepsilon) 
\ \left(\frac{1}{T_1}-\frac{1}{T_2}\right)
\exp\left[-\left(\frac{1}{T_1}-\frac{1}{T_2}\right)\varepsilon\right]
\eeq
One identifies that this is, up to the inverse temperature factor, 
merely the canonical average over $D$, leading 
to the desired result \Eq{Einstein}. 

\ \\

\twocolumngrid

{\bf Calculating the diffusion coefficient.-- }
The power spectrum of $W$ due to the evolution 
that is generated by $\mathcal{H}_0$ can be 
calculated from its matrix elements as follows:
\beq
\tilde{S}(\omega) &=& \mbox{FourierTransform} \  {\langle W(t)W(0)\rangle}
\\ \nonumber
&=& \sum_n p_n \sum_{m}|W_{m,n}|^2\ 2\pi\delta(\omega - (\varepsilon_m{-}\varepsilon_n))
\eeq
Here $p_n$ are the occupation probabilities 
of the $\mathcal{H}_0$ eigenstates.  
In order to calculate $\tilde{S}(\omega)$ for 
a given micro-canonical energy $\varepsilon$ 
the practical procedure is to plot the smoothed 
value $\overline{|W|^2}$ of the squared elements $|W_{m,n}|^2$ 
as a function of $\varepsilon = E_n$
along the diagonal ${(E_m - E_n) = \omega}$.
Then it follows that 
\beq
\tilde{S}(\omega) \ \ = \ \ 2\pi g(\varepsilon) \overline{|W^2|}
\eeq
If multiplied by the strength of the driving $|K_d|^2$, 
one obtains the Fermi-golden-rule expression 
for the rate of transitions due to a monochromatic 
driving. As implied by the Kubo formula \Eq{KuboD}
the diffusion coefficient is given by \cite{rspEq}:  
\beq
D(\varepsilon) \ \ = \ \ \frac{\pi}{2} \ (K_d \Omega)^2 \ g(\varepsilon) \overline{|W^2|}  
\eeq
where $\overline{|W|^2}$ has implicit dependence 
on both $\varepsilon$ and $\Omega$ as explained above.

\newpage
{\bf Semiclassical form of the trimer Hamiltonian.-- }
In a semi-classical context one substitutes $a_i=\sqrt{\bm{n}_i}\eexp{i\varphi_i}$, 
and defines ${q_i=\varphi_i-\varphi_0}$. 
Dropping a constant that depends of the conserved total 
particle number $N$, the BHH takes the form
\beq \nonumber
\mathcal{H} = -K_0\sum_{i=1,2}[\bm{n}_0\bm{n}_i]^{1/2}\cos q_i
-U\left[\bm{n}_0\bm{n}_1{+}\bm{n}_0\bm{n}_2{+}\bm{n}_1\bm{n}_2\right]
\eeq
%
%
%
Expressing $\bm{n}_0=N{-}\bm{n}_1{-}\bm{n}_2$ we see that the BHH is the quantized version 
of two coupled degrees of freedoms, where the effect of the interaction 
term (a quadratic function  of $\bm{n}_1$ and $\bm{n}_2$), is characterized 
by the dimensionless parameter $u=NU/K_0$. 

\ \\

{\bf Level statistics and the Brody parameter.-- }
Given $N$ and $K$ and $U$, we find the eigen-energies 
of the Hamiltonian \Eq{e1}. In each small energy range 
we calculate the mean level spacing, 
and the distribution $P(S)$ of the normalized spacings. 
Then we fit it to the Brody distribution \cite{Brody81} 
\beq
P_q(S) \ \ = \ \ \alpha S^q\exp(-\beta S^{1+q})
\eeq
with $\alpha=(1+q)\beta$, and $\beta=\Gamma^{1+q}\left[(2+q)/1+q)\right]$. 
Here $\Gamma$ denotes the Euler gamma function. 
A Brody parameter value of $q=0$ indicates a Poissonian level-spacing 
distribution characteristic of the uncorrelated levels of integrable system. 
By contrast for $q=1$ we have the Wigner level-spacing distribution, 
that reflects the level repulsion in the case of a quantized chaotic system. 
Thus, by plotting $q$ as a function of $\varepsilon$ 
we can map the domain of chaotic motion, see \Fig{f1}b. 

For large $u$ the dynamics is quasi-integrable due to self-trapped 
motion. The dynamics is trivially integrable also in the other extreme 
of very small~$u$. While we have also studied this latter region, 
the results concerning the response to driving were similar 
to the self-trapping integrability, and this point is not explicitly 
discussed in order to avoid redundancy.

\begin{figure}[h!]
\includegraphics[width=\hsize]{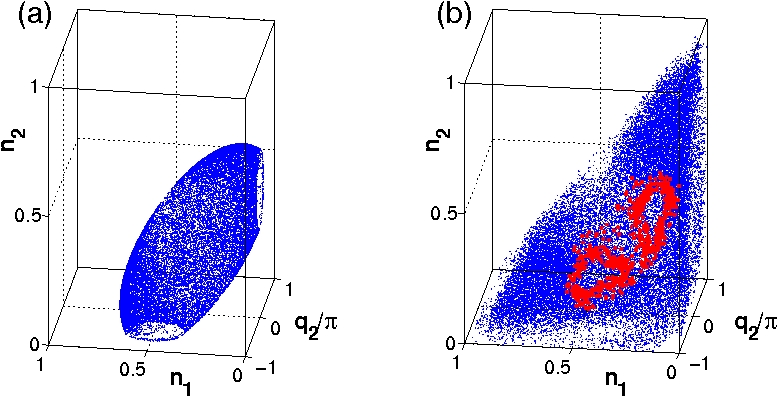}

\caption{
{\em Left panel:} 
The Poincare section of a representative 
classical chaotic trajectory.
The parameters are ${u=5}$ and ${\varepsilon=0.3}$.
The coordinates ${q_i=\varphi_i-\varphi_0}$ 
are conjugate to $n_i$, with ${i=1,2}$, 
and the section is at ${(q_1-q_2)=\pi/2}$. 
{\em Left panel:}
The square labeled trajectory of \Fig{f5}. 
is illustrated using the same coordinates as in the left panel. 
The points along the trajectory that have low energy (${\varepsilon<0.2}$)  
are highlighted by dark color: they reside within a sticky region. 
}

\label{f6}
\end{figure}

{\bf Poincare sections.-- }
The left panel of \Fig{f6} displays the Poincare section 
of a representative classical chaotic trajectory.
The right panel use the same coordinates for plotting 
a trajectory of the driven system. The sticky part of 
the trajectory is highlighted.

\clearpage
\end{document}